\documentclass[10pt]{article}
\usepackage[OE]{express}

\begin{document}

\title{Effect of scattering on coherent anti-Stokes Raman scattering (CARS) signals}

\author{Janaka~C.~Ranasinghesagara\authormark{1,2}, Giuseppe~de~Vito\authormark{3,4,5}, Vincenzo~Piazza\authormark{4}, Eric~O.~Potma\authormark{1,3}, and Vasan~Venugopalan\authormark{1,2,*}}

\address{\authormark{1}Beckman Laser Institute, University of California, Irvine,  California 92697, USA\\ \authormark{2}Department of Chemical Engineering and Materials Science, University of California, Irvine,  California 92697, USA\\
\authormark{3}Department of Chemistry, University of California, Irvine,  California 92697, USA\\
\authormark{4}Center for Nanotechnology Innovation @NEST, Istituto Italiano di Tecnologia, Piazza San Silvestro 12, I-56127, Pisa, Italy\\
\authormark{5}NEST, Scuola Normale Superiore, Piazza San Silvestro 12, I-56127 Pisa, Italy\\
}

\email{\authormark{*}vvenugop@uci.edu} %% email address is required



\begin{abstract}
We develop a computational framework to examine the factors responsible for scattering-induced distortions of coherent anti-Stokes Raman scattering (CARS) signals in turbid samples. We apply the Huygens-Fresnel Wave-based Electric Field Superposition (HF-WEFS) method combined with the radiating dipole approximation to compute the effects of scattering-induced distortions of focal excitation fields on the far-field CARS signal. We analyze the effect of spherical scatterers, placed in the vicinity of the focal volume, on the CARS signal emitted by different objects (2$\mu$m~diameter solid sphere, 2$\mu$m~diameter myelin cylinder and 2$\mu$m~diameter myelin tube). We find that distortions in the CARS signals arise not only from attenuation of the focal field but also from scattering-induced changes in the spatial phase that modifies the angular distribution of the CARS emission. Our simulations further show that CARS signal attenuation can be minimized by using a high numerical aperture condenser. Moreover, unlike the CARS intensity image, CARS images formed by taking the ratio of CARS signals obtained using $x\/$- and $y\/$-polarized input fields is relatively insensitive to the effects of spherical scatterers. Our computational framework provide a mechanistic approach to characterizing scattering-induced distortions in coherent imaging of turbid media and may inspire bottom-up approaches for adaptive optical methods for image correction.
\end{abstract}

%For a complete list of OCIS codes, visit: https://www.osapublishing.org/oe/submit/ocis/

%%%%%%%%%%%%%%%%%%%%%%% References %%%%%%%%%%%%%%%%%%%%%%%%%

%%%%%%%%%%%%%%%%%%%%%%%%%%  body  %%%%%%%%%%%%%%%%%%%%%%%%%%
\section{Introduction}
Coherent anti-Stokes Raman scattering (CARS) microscopy is a nonlinear, label-free imaging technique that has matured into a reliable tool for visualizing lipids, proteins and other endogenous compounds in biological tissues and cells based on their spatially-dependent third-order polarization \cite{Cheng2013, Chung2013, Zhang2015}. In the CARS process, a pair of incoming beams (named "pump" and "Stokes") are exploited to coherently and resonantly excite selected vibrational levels of a population of molecules. To this end, the beam frequencies are chosen so that their difference matches a vibrational frequency of the oscillating dipoles of interest. As a consequence of the interaction of the vibrationally excited molecule with a third photon, the nonlinear polarization radiates through emission of a fourth photon: the CARS signal~\cite{maker1965study}.

CARS microscopy is most commonly executed in the point illumination mode, in which the signal is generated in the focal volume of a high numerical aperture lens. Similar to all forms of microscopy that rely on the formation of a tight focal spot, the CARS signal is sensitive to the characteristics of the three-dimensional focal volume. Distorted or aberrated focal fields generally compromise CARS signal generation and degrade the CARS signal~\cite{Cheng2013,Zhang2015}. In contrast to other nonlinear optical microscopy techniques, such as two-photon excited fluorescence (TPEF)~\cite{Helmchen2005}, CARS microscopy relies on the spatial phase of the excitation field and is particularly sensitive to wavefront distortions. As a result, the CARS emission is dictated by both the amplitude and the phase of the focal fields. Moreover, since the pump beam and the Stokes beam have different wavelengths, their focal fields may exhibit different aberration characteristics.

The heterogeneity of biological samples, which results from structures of variable size and effective refractive index, modifies the propagation of focused optical wavefronts resulting in distorted focal volumes~\cite{Mosk2012}. The scattering-induced modification of the focal volume distribution is the primary factor for the deterioration of CARS signals at greater sample depths in turbid samples and results in attenuated signals, reduced contrast, and degraded resolution~\cite{DeAguiar2015}. While these effects may be less pronounced in thin samples such as cell cultures, refractive index variations still affect the focal volume and can alter the CARS radiation profiles, leading to signal loss or unaccounted image artifacts.

The deleterious effects of light scattering in coherent imaging methods can be mitigated by shaping the excitation optical wavefront to compensate for the anticipated scattering-induced wavefront distortions~\cite{Vellekoop2007,Yaqoob2008,Aguiar2016}. Such adaptive optics approaches offer the possibility to restore signal levels and retrieve high resolution images in turbid media~\cite{Mosk2012}. In the context of linear optical microscopy, wavefront shaping techniques have been used to almost completely counteract the effects of light scattering, or to leverage scattering in the medium to achieve image resolution surpassing that obtained in non-scattering samples~\cite{Mosk2012}. In recently published work, Judkewitz and co-workers~\cite{Papadopoulos2016} used scattered electric field point spread function as a guidance to compensate the effect of scattering. Such adaptive optics method may not work when transmission signals acquired from reference and scattered  beams lack sufficient correlation. Adaptive optics approaches have also been applied to CARS microscopy, by using the maximization of the CARS intensity as an objective function to optimize the shaping of the excitation wavefront~\cite{Wright2007}. 

Virtually all adaptive optics approaches are based on empirical optimization of  experimentally accessible parameters, such as the signal intensity~\cite{Vellekoop2007,Yaqoob2008,Aguiar2016,Papadopoulos2016,Wright2007}. In this approach, the sample is considered a black box, which can be characterized by an effective transmission matrix that does not require a detailed understanding of the physical origin of the wavefront distortions. In many cases, such a strategy has proven to work well for counteracting scattering effects in linear optical microscopy applications. However, in nonlinear optical microscopy, there is evidence that maximizing signal intensity may not represent an appropriate optimization metric, resulting in the convergence to local extrema that correspond to focal shapes and positions that are markedly different from those obtained under non-scattering conditions~\cite{Katz2014}. This possibility underscores that a general strategy to manage the deleterious effect of light scattering effects must go beyond empirical optimization of signal intensities. This notion is particularly pertinent to CARS microscopy, where subtle amplitude and phase effects can have dramatic consequences for the observed signal intensities~\cite{Cheng2002a}. Instead of tackling the problem through an empirical black box approach, a fundamental understanding of the physics that gives rise to scattering artifacts in CARS is imperative. In this regard, a bottom-up, computational approach, that considers how wavefront aberrations affect CARS imaging, may provide the insights necessary to devise experimental approaches for recording CARS images devoid of scattering artifacts.
 
Such a detailed, fundamental understanding of linear scattering effects in coherent Raman scattering does not currently exist. Several model-based approaches have been used to investigate the effect of light scattering on the generation of coherent Raman signals in scattering media~\cite{Zhu2013,Hokr2014}. These include the use of Monte Carlo methods to simulate Raman scattering in turbid samples~\cite{Hokr2014}. However, Monte Carlo simulations are unable to rigorously model diffraction or properly account for the amplitude and phase characteristics of propagating fields. These deficiencies prevent Monte Carlo simulations from accurately modeling spatial coherence, which is a critical determinant for the generation of coherent nonlinear optical signals. While full-field simulations can be conducted using finite-difference time domain (FDTD) methods to study the effect of scatterer size and orientation on near-field CARS signals~\cite{Lin2009,Lin2010,vanderKolk2016}, they are impractical for extensive parametric studies due to the substantial computational cost~\cite{dunn2009OE}.

In this work, we aim to take several important steps toward building a fundamental, real-space picture of how linear scattering affects experimental observables in CARS microscopy. Recently we introduced a new efficient method to compute focal field distortions produced by scattering particles using Huygens-Fresnel wavelet propagation~\cite{Prahl2010a} and field superposition methods~\cite{Ranasinghesagara2014}. This Huygens-Fresnel Wave-based Electric-Field Superposition (HF-WEFS) approach provides accurate focal field predictions in the presence of single or multiple scatterers with arbitrary size, spatial configuration, density and orientation. Here, we apply a computational framework that employs HF-WEFS to examine CARS signal generation and far-field detection in the presence of scattering. Our framework first employs the HF-WEFS method to compute scattering-induced focal volume distortions of both the pump and Stokes beams. Next, we determine the CARS signal generation by computing the spatially dependent third-order dielectric polarization density produced by the pump and Stokes fields. Finally, we use the radiating dipole approximation~\cite{Cheng2013,Krishnamachari2007} to compute the CARS signal as measured by a far-field detector. This approach enables the simulation of the far-field CARS signal with pump and Stokes beams of arbitrary polarization state, spatial distribution, illumination and detection numerical aperture, scatterer configuration, and scatterer shape. 

As test samples, we simulate an isotropic solid sphere, a myelin cylinder and a myelin tubular structure. Myelin is a biological structure that envelopes a subgroup of nerve fibers in the gnathostomata and functions to increase nerve impulse conduction efficiency. We chose to simulate myelin due to its biological relevance, morphology and molecular characteristics that make it suitable for CARS imaging.  CARS microscopy is frequently employed in myelin imaging, thanks to the strong CARS signal obtained by targeting its extremely abundant CH$_{2}$ bonds. Consequently, myelin has been studied using CARS imaging under normal physiologic~\cite{wang2005coherent,canta2016age} and pathologic~\cite{fu2011paranodal,imitola2011multimodal,deVito2016rp,shi2011longitudinal} conditions. 

\section{Methods}
\begin{figure}[ht]
	\begin{center}
		\includegraphics[width=0.65\columnwidth]{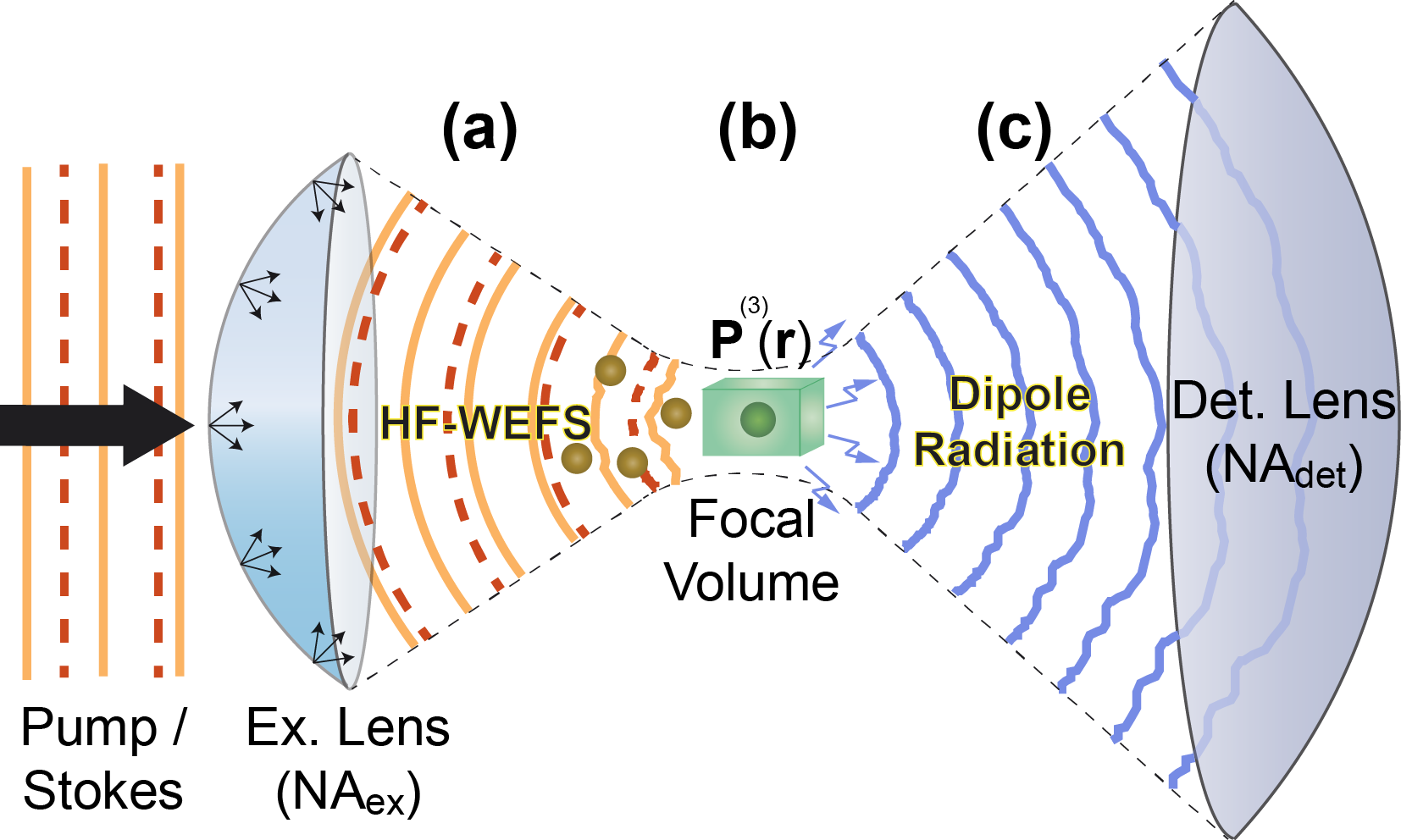}
	\end{center}
	\caption{Illustration of focused beam propagation, CARS signal generation in the focal volume and signal emission. (a) The HF plane waves of pump and Stokes beams propagate separately in a medium with scatterers. (b) The spatially dependent polarization is computed in the focal volume. (c) Dipole radiation and the far-field detection. The lens are geometrically represented by reference spherical surfaces. Numerical aperture of the excitation and detection lenses are NA$_{\mathrm{ex}}$ and NA$_{\mathrm{det}}$, respectively.
		\label{fig:Figure1}}
\end{figure}

Our framework deconstructs the process of CARS excitation, emission and detection into three sequential computations: (a) focused beam propagation in a scattering medium, (b)~production of a nonlinear polarization field within the focal volume, and (c) far-field dipole radiation. A schematic of these components is shown in Fig.~\ref{fig:Figure1}. We use the HF-WEFS method to rigorously model the focal fields generated by the propagation of pump and Stokes beams.

Once the focal fields have been computed, we compute the third-order dielectric polarization density, $\mathbf{P}^{(3)}(\mathbf{r})$ produced by the incident pump and Stokes electric field distributions within the focal volume based on the nonlinear susceptibility of the medium. The emission that follows from $\mathbf{P}^{(3)}(\mathbf{r})$ is then modeled as a collection of radiating dipoles in focus, which couple to, and is detected in, the far-field~\cite{Cheng2013,Krishnamachari2007}.

We detail each of the processes represented in Fig.~\ref{fig:Figure1} in the following subsections. 

\subsection{Focus Beam Propagation}
We consider monochromatic pump and Stokes beams incident upon an aplanatic lens, and propagating independently towards the focal volume. In this study, we use the fundamental Hermite-Gaussian spatial mode (HG00) for both pump and Stokes beams. The electric field amplitude distribution of a Gaussian beam at the plane of an aplanatic lens can be expressed as~\cite{Novotny2006}:

\begin{equation}
|{\mathbf E}_{\mathrm{inc}}(x,y)| =
E_0\,\mathrm{exp}[-(x^2+y^2)/\omega_0^2]
\label{eq:EqGauss},
\end{equation}

\noindent where $E_0=1$ and $\omega_0\/$ is the radius of the Gaussian beam at which the electric field amplitude falls to $1/\mathrm{e}$ of the maximum axial value. The aplanatic lens system can be geometrically represented by a reference spherical surface that has a center at the origin~\cite{Novotny2006,Ranasinghesagara2014}. The HF-WEFS method considers forward propagation of Huygens-Fresnel spherical waves from the reference spherical surface. We determine the propagation origin of each HF spherical wave at the lens surface by generating a set of uniformly distributed points on the reference surface~\cite{Koay2011}. Each spherical wave is represented by the summation of outward propagating Huygens-Fresnel plane wavelets~(HF wavelets)~\cite{Ranasinghesagara2014,Richards1959}. In the absence of linear scattering in the space between the lens surface and the focal region, this Huygens-Fresnel description accurately reproduces the three-dimensional, diffracted-limited focal volume as predicted by diffraction theory~\cite{Richards1959}. The amplitude of an HF wavelet at each radiating point is given by $|\mathbf{E}_{\mathrm{inc}}(x,y)|$. The parallel and perpendicular electric field components $(E_{\parallel},E_{\perp})$ of the HF wavelet at the spherical reference surface are given by~\cite{Ranasinghesagara2014}:

\begin{equation}
\left(\begin{array}{c} 
E_{\parallel}\\ E_{\perp}
\end{array}
\right) =
|\mathbf{E}_{\mathrm{inc}}(x,y)|
\left(
\begin{array}{cc}
\cos\phi & \sin\phi \\
-\sin\phi & \cos\phi
\end{array} 
\right)  
\left(
\begin{array}{c} 
\mathbf{JV}
\end{array}\right)  
\sqrt\frac{n_{\mathrm{inc}}}{n}(\cos\theta)^{\frac{1}{2}},
\label{eq:EqHFIncident}
\end{equation}

\noindent where $\mathbf{JV}$ is the Jones vector that describes the polarization of light, and $n_{\mathrm{inc}}$ and $n$ are the refractive indices of the medium before and after the lens. $\phi$ and $\theta$ are azimuthal and polar angles of the HF plane wavelet with respect to the global coordinate system. The unscattered electric field components ($E_{\parallel}^{\mathrm{unscat}}, E_{\perp}^{\mathrm{unscat}}$) at a distance $d$ from the point of emission can be expressed as:

\begin{equation}
\left(\begin{array}{c} 
E_{\parallel}^{\mathrm{unscat}}\\ E_{\perp}^{\mathrm{unscat}}
\end{array}
\right)=\left(
\begin{array}{c} 
E_{\parallel}\\ E_{\perp}
\end{array}\right)
\exp(-ikd),
\label{eq:EqUnscatteredField}
\end{equation}
where $k\/$ is the wave number $=2\pi/\lambda\/$.

When scatterers are present in the medium, we consider each scatterer sequentially and account for all possible HF plane wavelets that may interact with it. In this study, we select spherical scattering particles, for which full-amplitude scattering matrices can be readily obtained using Lorenz-Mie theory~\cite{VandeHulst1957}. For a scatterer located at point $D$, the parallel and perpendicular polarization components of the scattered electric field for a specific polar angle $\theta_\mathrm{s}$ and distance from the scatterer $r_\mathrm{s}$ can be expressed as \cite{Ranasinghesagara2014}:

\begin{equation} 
\left(\begin{array}{c} 
E_{\parallel}^{\mathrm{scat}}\\ E_{\perp}^{\mathrm{scat}}
\end{array}
\right) = 
\frac{1}{kr_\mathrm{s}}
\left( 
\begin{array}{cc} 
S_2(r_\mathrm{s},\theta_\mathrm{s}) &  0\\
0   &  S_1(r_\mathrm{s},\theta_\mathrm{s})
\end{array}\right)
\left(
\begin{array}{cc}
\cos\phi_\mathrm{s} & \sin\phi_\mathrm{s} \\
-\sin\phi_\mathrm{s} & \cos\phi_\mathrm{s}
\end{array}
\right) 
\left(
\begin{array}{c} 
E_{\parallel D}\\ E_{\perp D}
\end{array}\right),
\label{eq:EqScatField}
\end{equation}

\noindent where $E_{\parallel D}$ and $E_{\perp D}$ are the parallel and perpendicular incident electric field components at point $D$. The unscatterted and scattered fields can be superposed to obtain the total field at any location~\cite{VandeHulst1957}. The parallel and perpendicular electric field components calculated in Eqs.~\ref{eq:EqUnscatteredField} and \ref{eq:EqScatField} are transformed into $x$, $y$, and $z$ components before superposition~\cite{Ranasinghesagara2014}. The components of the total electric field at a location $\mathbf{r}$, $\mathbf{E}(\mathbf{r})\/$, can be computed as: 

\begin{equation}
\left(\begin{array}{c} 
E_{x}(\mathbf{r})\\ E_{y}(\mathbf{r}) \\ E_{z}(\mathbf{r})
\end{array}
\right)=
\left(\begin{array}{c} 
E_{x}^{\mathrm{unscat}}(\mathbf{r})+E_{x}^{\mathrm{scat}}(\mathbf{r}) \\
E_{y}^{\mathrm{unscat}}(\mathbf{r})+E_{y}^{\mathrm{scat}}(\mathbf{r}) \\ 
E_{z}^{\mathrm{unscat}}(\mathbf{r})+E_{z}^{\mathrm{scat}}(\mathbf{r})
\end{array}
\right),
\label{eq:EqTotal}
\end{equation}

Equations \ref{eq:EqGauss}--\ref{eq:EqTotal} are used to propagate the pump beam $\mathbf{E}_{\mathrm{p}}(\mathbf{r})\/$ and Stokes beam $\mathbf{E}_{\mathrm{S}}(\mathbf{r})\/$ in a scattering medium to obtain their $x$, $y$, and $z$ components of the electric field in the focal volume.

%%%%%%%%%%%%%%%%%%%%%%%%%%%%%%%%%%%
\subsection{Polarization Signal Computation}
In CARS microscopy, the $i^{\mathrm{th}}\/$ component of the spatially-dependent third-order dielectric polarization density induced at location $\mathbf{r}$ by the pump electric field and Stokes electric field is computed from: 

\begin{equation}
{P}^{(3)}_{i}(\mathbf{r}) =
\sum_{j,k,l}\chi_{ijkl}^{(3)}(\mathbf{r})
{E}_{\mathrm{p}j}(\mathbf{r})
{E}_{\mathrm{p}k}(\mathbf{r})
{E}_{\mathrm{S}l}^{*}(\mathbf{r})
\label{eq:EqPolarization},
\end{equation}

\noindent where $i=(x,y,z)\/$ and $\chi_{ijkl}^{(3)}(\mathbf{r})$ is the third-order non-linear susceptibility tensor of the objects or media. ${E}_{\mathrm{p}j}(\mathbf{r})$ and ${E}_{\mathrm{p}k}(\mathbf{r})$ are electric field components of the pump beam and ${E}_{\mathrm{S}l}^{*}(\mathbf{r})$ is conjugate electric field components of the Stokes beam at location $\mathbf{r}$.

In this study, we consider the third-order nonlinear susceptibility tensor of spherical objects and cylindrical and tubular myelin structures placed within the focal volume. The nonlinear susceptibility is a tensor of rank 4, with 81 elements. The number of nonzero and independent elements depends on the spatial symmetry of the sample object. We assume the spherical objects to be uniform and isotropic, which results in 21 nonzero tensor elements, of which only four are independent ($\chi^{(3)}_{xxxx}=\chi^{(3)}_{xxyy}+\chi^{(3)}_{xyxy}+\chi^{(3)}_{xyyx}$)~\cite{Boyd2003}. For the cylindrical and tubular myelin structures, we employed published tensor element values that were experimentally determined for myelin sheaths~\cite{deVito2012}. Although different from the isotropic case, the nonlinear susceptibility of myelin sheaths is also described by 21 nonzero elements, with four independent tensor elements~\cite{Mazely1987,Belanger2009,deVito2012,deVito2014rp}. Because myelin layers are organized in concentric cylinders, their constituent molecules are rotated with respect to the laboratory frame depending on the location in the myelin structure. To model the measured response in the laboratory frame, the molecular nonlinear susceptibility is rotated with the proper Euler angles to find the overall CARS response of the system~\cite{deVito2014rp}.

%%%%%%%%%%%%%%%%%%%%%%%%%%%%%%%%%%%
\subsection{Far-field Dipole Radiation}
Once the nonlinear polarization is determined within the focal volume, the resulting far-field CARS emission can be modeled using an ensemble of radiating dipoles~\cite{Cheng2013,Krishnamachari2007}. For this purpose, each volume element in the vicinity of the focus is considered a point dipole. The magnitude of the dipole is given by Eq.~\ref{eq:EqPolarization}. Each dipole radiates, and the resulting electric field is detected in the far field. The total amplitude of the electric field, $\mathbf{E}_\mathrm{C}(\mathbf{R})$, at a far field location $\mathbf{R}$ is the sum of the amplitude contributions from all point dipoles emanating from $\mathbf{r}$~\cite{Jackson1975,Novotny2006}:  

\begin{equation}
\mathbf{E}_\mathrm{C}(\mathbf{R;r}) =
\int_{\mathcal{V}}^{}\frac{e^{ik_\mathrm{C}|\mathbf{R-r}|}}{4\pi|\mathbf{R-r}|^3}
[(\mathbf{R-r})\times\mathbf{P}^{(3)}(\mathbf{r})]
\times(\mathbf{R-r})\,\mathrm{d}\mathcal{V}
\label{eq:EqDipole},
\end{equation}

\noindent where $k_\mathrm{C}=2\pi/\lambda_\mathrm{C}$, $\lambda_\mathrm{C}$ is the CARS wavelength in the medium, and $\mathcal{V}$ is the excitation volume. To calculate angular resolved CARS radiation patterns shown in Figs.~3 and 4, we compute $|\mathbf{E}_\mathrm{C}(\mathbf{R})|^2$ by making use of Eq.~(\ref{eq:EqDipole}). The total CARS signal intensity $I_\mathrm{C}$ captured by the far-field lens with an acceptance angle of $\alpha_{max}$ can be written as~\cite{Cheng2002a}

\begin{equation}
I_\mathrm{C} \propto
\int_{\theta=0}^{\alpha_{max}}\int_{\phi=0}^{2\pi}
|\mathbf{E}_\mathrm{C}(\mathbf{R})|^2|\mathbf{R}|^2 
\,\sin\theta\, \mathrm{d}\phi\,\mathrm{d}\theta
\label{eq:EqIntensity}
\end{equation}

To obtain the CARS intensity as a function of the $y$--$z$ grid (Fig. 5) and to simulate CARS images (Figs.~6 and 7), we compute the total CARS intensity using Eq.~\ref{eq:EqIntensity}.

%%%%%%%%%%%%%%%%%%%%%%%%%%%%%%%%%%%%%
\subsection{Numerical Simulation}
In this study, the wavelengths of pump and Stokes beams are selected as $\lambda = 800\/$~nm and 1064~nm, respectively. We consider HG00 beams with filling factors (~=~$\omega_0/f$NA$_{\mathrm{ex}}$) equal to unity\cite{Novotny2006}, where $f$ is the focal length of the lens. We consider $\left(n/n_{inc}\right) =1$ and compute the excitation within a  3\,$\mu$m\,$\times$\,3\,$\mu$m\,$\times$\,6\,$\mu$m volume centered about the focal point. This volume is subdivided into a three-dimensional grid with 50~nm cubic voxels. We compute the distorted pump and Stokes electric fields at each grid point separately using Eqs.~\ref{eq:EqGauss}--\ref{eq:EqTotal}. 

We consider the CARS imaging of three separate objects: 2~$\mu$m~diameter sphere, 2~$\mu$m~diameter myelin cylinder and 2~$\mu$m~diameter myelin tube. The myelin tube has wall thickness of 250~nm and is centered or offset from the optical axis. The refractive indices of the medium and the scatterers are 1.33 and 1.49, respectively. Even while the CARS active objects have  different refractive indices, we assume them to be index matched when modeling light propagation. The $\chi^{(3)}$ of each object is considered as non-resonant, i.e., we ignore tentative phase effects due to the presence of spectral resonances. The values of the nonlinear susceptibility tensor elements of the objects are obtained as described above. The $\chi^{(3)}_{ijkl}$ of the surrounding medium, including the empty center portion of the tube, is set to zero. The $x,y,z$ components of $\mathbf{P}^{(3)}(\mathbf{r})$ are computed using Eq.~\ref{eq:EqPolarization}, with $\chi^{(3)}_{ijkl}$ and the electric field distribution of pump and Stokes beams as inputs. After calculating $\mathbf{P}^{(3)}(\mathbf{r})$ in the volume element of each grid point, the far-field amplitude is computed using Eq.~\ref{eq:EqDipole}. Computation of the CARS far-field emission is accomplished by placing a hemispherical detector in the far-field. The total CARS intensity is computed by integrating the far-field CARS radiation pattern over the detector acceptance angle, as in Eq.~\ref{eq:EqIntensity}. We consider detection with acceptance angles of 71.8$^\circ$~(NA$_{\mathrm{det}}$~=~0.95) and 33.4$^\circ$~(NA$_{\mathrm{det}}$~=~0.55). 

\begin{figure}[ht!]
	\begin{center}
		\includegraphics[width=0.86\columnwidth]{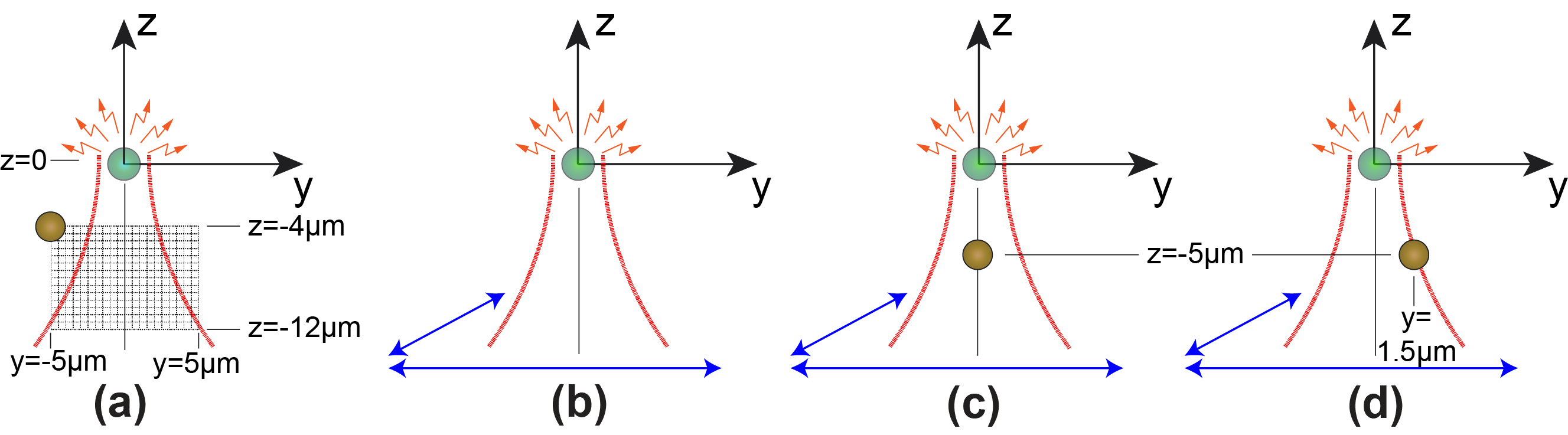}
	\end{center}
	\caption{Simulation Setup. (a) A 2~$\mu$m diameter spherical scatterer (gold) is placed at different locations of $y$-$z$ grid $(x=0)\/$ to obtain its effect on the CARS intensity. (b)--(d) The lens system is scanned in the $x$-$y$ plane while keeping the object (green) and the spherical scatterer stationary. We consider CARS imaging in a (b) non-scattering medium; and in systems containing a spherical scatterer placed at (c) $(x,y,z)=(0,0,-5)\,\mu$m and (d) $(x,y,z)=(0,1.5,-5)\,\mu$m.
		\label{fig:Figure2}}
\end{figure}

We examine the CARS signal under scattering and non-scattering conditions for different excitation numerical aperture (NA$_{\mathrm{ex}}$~=~0.825 and 0.55). Figure \ref{fig:Figure2} depicts the various simulation geometries. In Fig.~\ref{fig:Figure2}(a) we depict the effect of scatterer locations within the $y$-$z$ grid on the far-field CARS intensity. The $y$-$z$ grid has an overall dimension of $(y,z) = 10\,\mu\/$m$\times 8\,\mu\/$m  with 0.5~$\mu$m spacing.  In Fig.~\ref{fig:Figure2}(b) we depict the generation of CARS images using point illumination without scattering as references. In Figs.~\ref{fig:Figure2}(c) and (d), we depict two cases used to examine the effects of a discrete scatterer on CARS imaging. Figure \ref{fig:Figure2}(c) considers the effect of a 2~$\mu$m diameter scatterer placed along the optical axis 5~$\mu$m below the focal plane. Figure \ref{fig:Figure2}(d) considers the same scatterer placed at the same depth but offset 1.5~$\mu$m to the right of the optical axis. We consider CARS images generated using $x\/$-polarized and $y\/$-polarized light for both pump and Stokes beams separately.  

%%%%%%%%%%%%%%%%%%%%% Results and Discussion  %%%%%%%%%%%%%%%%%%%%%%
\section{Results and Discussion}

%%%%%%%%%%%%%% A %%%%%%%%%%%%%%%%%%
\subsection{CARS radiation profiles with a scatterer}
\begin{figure}[ht!]
	\begin{center}
		\includegraphics[width=0.88\textwidth]{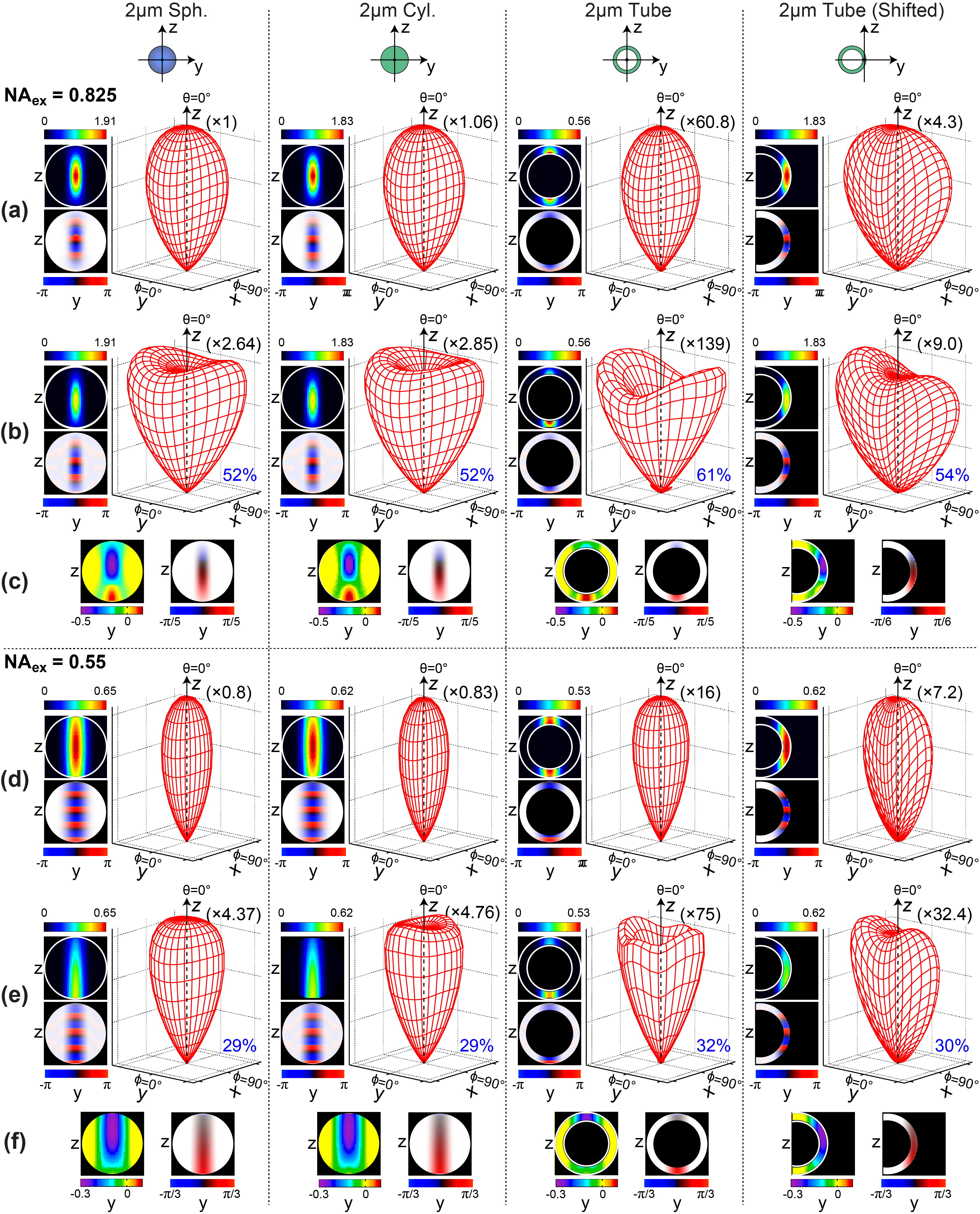}
     \end{center}
	\caption{Far-field CARS radiation patterns (from L to R) from a 2~$\mu$m diameter solid sphere, 2~$\mu$m diameter myelin cylinder, and 2~$\mu$m diameter myelin tube (centered and shifted by 0.875~$\mu$m left of the optical axis) in a (a,d) non-scattering medium and (b,e) medium with scatterer placed at $(x,y,z)=(0,0,-5)\mu$m. NA$_{\mathrm{ex}}$~=~0.825~in~(a,b,c) and NA$_{\mathrm{ex}}$~=~0.55~in~(d,e,f). Insets to the left of each radiation pattern show $y\/$-$z\/$ cross-sections of the amplitude (upper) and phase (lower) of $\mathbf{P}^{(3)}(\mathbf{r})$. Insets in rows (c) and (f) show the amplitude (left) and phase (right) \textit{differences} of (b) and (e) relative to the corresponding non-scattering cases, (a) and (d), respectively. Each inset spans 2$\mu\/$m~$\times$~2$\mu\/$m. Each radiation profile was multiplied by the number in the bracket to provide same maximum radiance. The percentages in (b) and (e) indicate the CARS intensity relative to the corresponding non-scattering case. Detection numerical aperture is fixed at NA$_{\mathrm{det}}$~=~0.95. 
		\label{fig:Figure3}}
\end{figure}

We first consider the CARS radiation profiles resulting from pump and Stokes beams in the absence and presence of scattering objects. Figure~\ref{fig:Figure3}(a) provides far-field CARS radiation patterns for different objects located at the focus in a medium without scattering. In these computations, the excitation fields are focused by a microscope objective with NA$_{\mathrm{ex}}=0.825$ and the far-field radiation is detected using a lens with NA$_{\mathrm{det}}=0.95$. The CARS emission intensity is shown as a function of the collection angle in the far-field. For comparison, each radiation profile is multiplied by a factor shown in brackets to provide plots have the same maximum radiance relative to the non-scattering case of the 2~$\mu$m spherical object. The inset of each panel shows the amplitude and phase cross sections ($y$-$z$ plane, $x$=0) of $\mathbf{P}^{(3)}(\mathbf{r})$ in the focal volume. Displays of the phase cross-section are masked with the amplitude distribution to emphasize the regions of the focal volume that contribute most significantly to the CARS emission.  

In the non-scattering case, the CARS emission is highly forward directed and results from the phase-matching of the CARS radiation along the optical axis~\cite{Cheng2002a}. This situation changes when a scatterer is introduced in the vicinity of the focal excitation volume. In Fig.~\ref{fig:Figure3}(b), we show CARS radiation profiles for these same objects in cases where a scatterer is positioned along the optical axis  5\,$\mu\/$m below the focal plane. The insets show profiles of the amplitude and phase \textit{differences} of $\mathbf{P}^{(3)}(\mathbf{r})$ in the focal volume relative to the non-scattering case. The wavefront aberrations produced by the scattering object result in a nominal shift of the maximum amplitude of the polarization density to positions just below the focal plane. The scattering also distorts the phase profile of the induced polarization. Along the optical axis, scattering of the excitation fields introduces an extra phase shift in the nonlinear polarization approaching ${\pi}/{3}$ across the focal volume. This additional phase shift is responsible for the reduced intensity and modified angular distribution of the CARS radiation profiles. When the object is centered about the optical axis, the forward directed CARS signal is depleted significantly, whereas the off-axis radiation is more prominent. This is also observed in the shifted myelin tube (fourth column of Fig.~\ref{fig:Figure3}(b)) where uneven amplitude and phase profiles within the tube contribute to an asymmetric CARS emission profile. These results clearly illustrate that the presence of a scattering particle not only modifies the overall amplitude of the nonlinear polarization, but also the spatial phase distribution. The lobed radiation pattern results from a scattering-induced phase shift along the optical axis, as can be inferred from the spatial phase profiles of the nonlinear polarization. This observation highlights the need to consider both the amplitude and phase of the excitation fields to properly account for the interference effects that occur within the focal volume.

In Fig.~\ref{fig:Figure3}(c) we show how these CARS radiation patterns are altered when using excitation illumination with a reduced numerical aperture (NA$_{\mathrm{ex}}=0.55$) in a non-scattering medium, while keeping the detection NA unchanged (NA$_{\mathrm{det}}=0.95$). The smaller illumination NA$_{\mathrm{ex}}$ introduces a narrower range of spatial frequencies into the sample and results in a broader and more elongated focal excitation volume. The longer interaction volume provides a more directional, phase-matched CARS signal along the $z\/$-axis. Figure \ref{fig:Figure3}(d) displays the CARS radiation profiles in the presence of the scatterer. In contrast to the results of Fig.~\ref{fig:Figure3}(b), when using the smaller illumination NA the CARS signals from the solid sphere and cylinder remain highly directional along the optical axis with much smaller changes in the radiation pattern. This shows that the CARS emission of solid objects are less sensitive to phase aberrations carried by the smaller spatial frequencies associated with the lower NA of illumination. By contrast, CARS emission from the hollow myelin tubes remain sensitive to scattering-induced phase changes carried by the lower spatial frequencies of the excitation light, resulting in more CARS radiation profiles that remain distorted. Also noteworthy is the lack of attenuation of the CARS signal in the case when the edge  of the myelin tube is at focus, as compared to the centered case.

The percentage values shown in Fig.~\ref{fig:Figure3}(b) and Fig.~\ref{fig:Figure3}(d) provide the total CARS signal intensity relative to the non-scattering case for each NA$_{\mathrm{ex}}$. As expected, the presence of the scatterer attenuates the CARS intensity for all objects. The attenuation relative to the non-scattering case is larger when illuminating the sample with the lower NA. However, the angular distribution of the CARS radiation remains very directional. 

\begin{figure}[ht!]
	\begin{center}
		\includegraphics[width=0.84\textwidth]{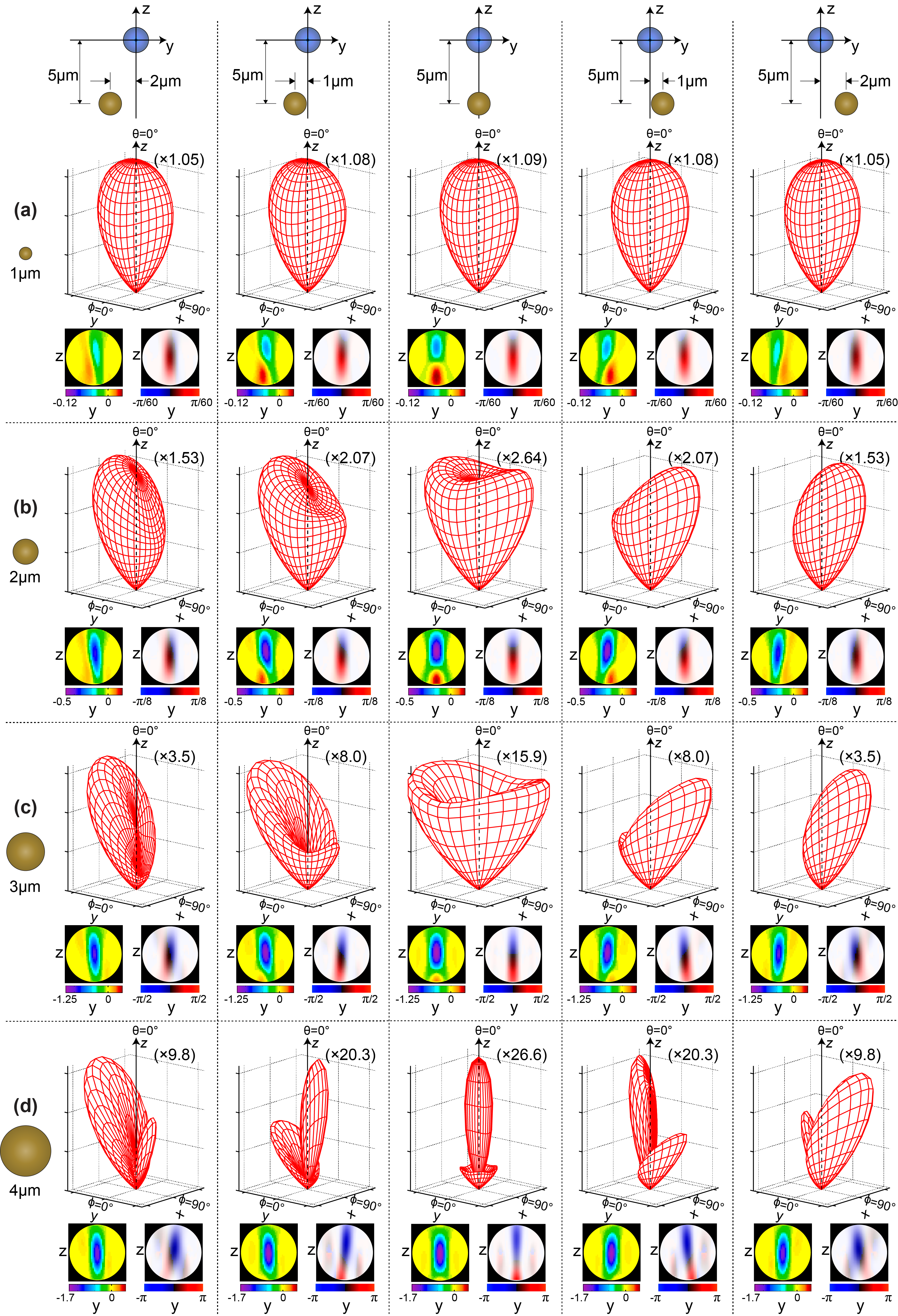}
	\end{center}
	\caption{The far-field CARS radiation patterns from a 2~$\mu$m diameter solid sphere (blue) located at the focal point in a medium with a single scatterer (gold) placed 5$\mu\/$m below the optical plane at $y\/$ locations of $y=-2,-1,0,1,2\,\mu\/$m as shown. The effect of scatterer size is shown for diameters of (a) 1\,$\mu$m, (b) 2\,$\mu$m, (c) 3\,$\mu$m, and (d) 4\,$\mu$m. Each radiation profile was multiplied by the number in the bracket to provide same maximum radiance. Insets to the bottom of each radiation pattern show $y\/$-$z\/$ cross-sections of  amplitude \textit{difference} (left) and phase \textit{difference} (right). Amplitude/phase \textit{differences} are calculated by subtracting amplitude/phase of $\mathbf{P}^{(3)}(\mathbf{r})$ induced in a non scattering medium. Excitation and detection numerical apertures are fixed at NA$_{\mathrm{ex}}$~=~0.825 and NA$_{\mathrm{det}}$~=~0.95.	
		\label{fig:Figure4}}
\end{figure}

In Fig.~\ref{fig:Figure4}, we show the effect of lateral particle position on the CARS radiation patterns for single scatterers placed at different positions along the $y\/$-axis, 5\,$\mu\/$m below the focal plane ($z\/=-5\,\mu\/$m). We consider scatterer locations of $y=-2,-1,0,1\/$ and $2\,\mu$m. The scatterer diameter is varied from 1 to 4\,$\mu$m (Figs.~\ref{fig:Figure4}(a)--(d)). For a 1$\mu\/$m scatterer diameter (Fig.~\ref{fig:Figure4}(a)), the distortion and attenuation of the far-field radiation profile is minimal. Larger scattering particles provide more substantial amplitude attenuation and phase distortion resulting in more pronounced variations in the CARS radiation profiles and overall signal attenuation. The largest attenuation and distorted radiation profiles are seen for the 4\,$\mu\/$m diameter scatterer (Fig.~\ref{fig:Figure4}(d)) because the scattering induced phase shift in the nonlinear polarization along the optical axis approaches 2$\pi$ (Fig.~\ref{fig:Figure4}(d)). The peak intensity along the optical axis is greatly affected for scatterer locations directly under the spherical object. Importantly, highly asymmetric radiation profiles are produced when the particle is displaced laterally from the optical axis. These asymmetric profiles result from spatial phase distortions carried by the nonlinear polarization in the focal volume and directly impact the CARS signal detection. We provide computed amplitude \textit{differences} and phase \textit{differences} of the nonlinear polarization relative to the non-scattering case. Figure \ref{fig:Figure4} demonstrates the impact of lateral scatterer position on the angular profile of the CARS emission.

%%%%%%%%%%%%%  B  %%%%%%%%%%%%%%%%%%%%%%
\subsection{Effect of scatterer position and detection numerical aperture on total CARS intensity}

\begin{figure}[ht!]
	\begin{center}
		\includegraphics[width=0.9\textwidth]{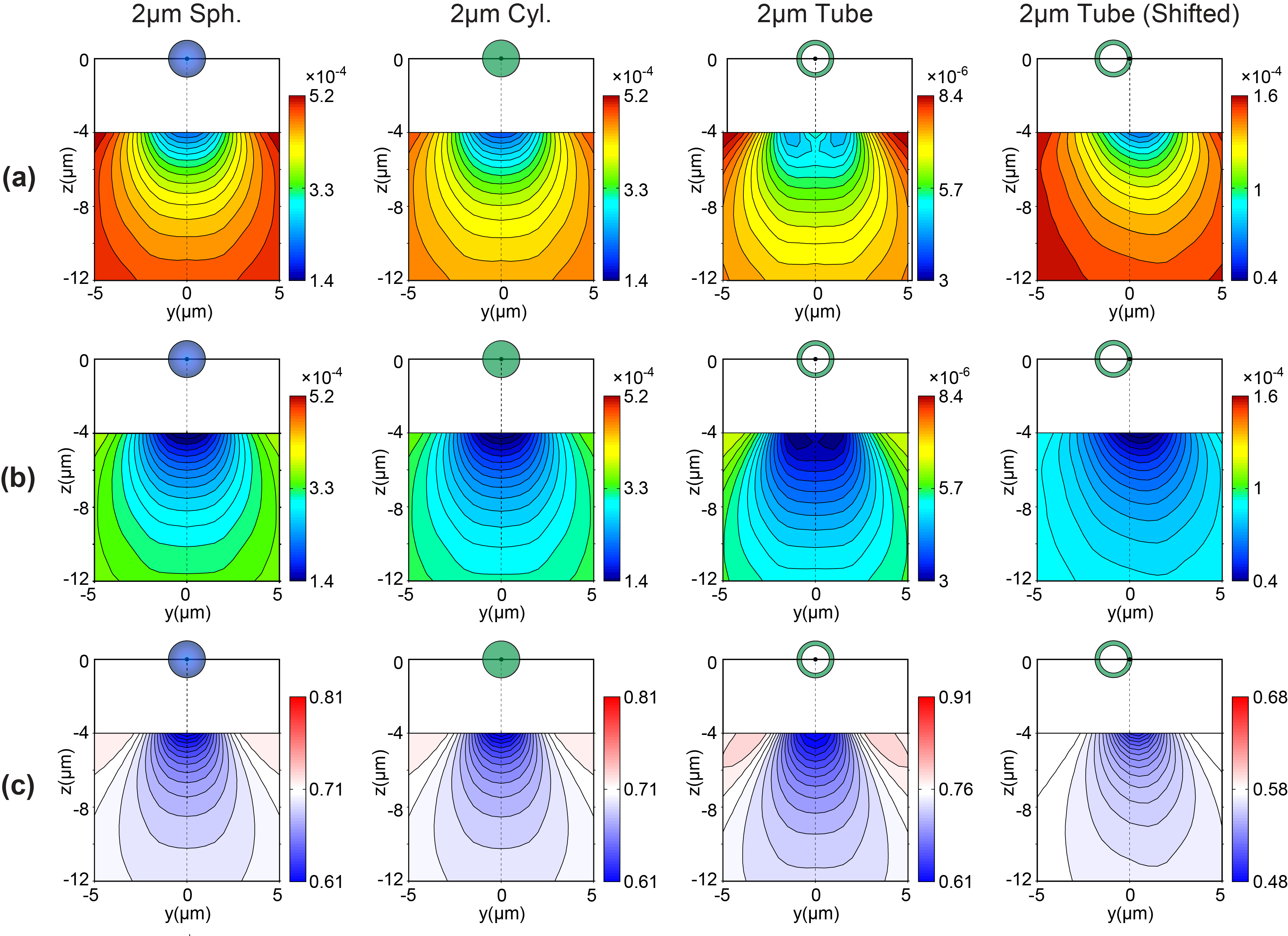}
	\end{center}
	\caption{The far-field CARS intensity as a function of the $y$-$z$ particle location grid. The object that is placed at the focus is 2~$\mu$m sphere, 2~$\mu$m myelin cylinder, and 2~$\mu$m myelin tube (centered and shifted by 0.875~$\mu$m left of the optical axis). Scatterer diameter is 2~$\mu$m. NA$_{\mathrm{det}}$ is (a) 0.95 and (b) 0.55. (c) Intensity ratio after dividing (b) by (a). The center value (white color) of the ratio color bar represents the ratio obtained for the non scattering medium. Excitation numerical aperture is fixed at NA$_{\mathrm{ex}}$~=~0.825.
		\label{fig:Figure5}}
\end{figure}

We now examine how the $y\/$-$z\/$ position of a single scattering particle impacts the total CARS intensity with NA$_{\mathrm{det}}$~=~0.95. Figure~\ref{fig:Figure5} shows the integrated CARS signal from several objects as a function of position of a 2~$\mu$m diameter spherical scatterer. The scatterer positions correspond to those in the $y$-$z$ grid shown in Fig.~\ref{fig:Figure2}(a). We start from $z = -4\,\mu\/$m to avoid overlap between the scattering particle and the focal volume under consideration. Figure \ref{fig:Figure5}(a) shows that for the case of imaging a 2~$\mu$m diameter solid sphere, myelin cylinder and myelin tube, the attenuation of the CARS signal is more prominent for scatterer positions more proximal to the focal volume. 

Figure \ref{fig:Figure5}(b) provides these same results for a reduced detection numerical aperture of NA$_{\mathrm{det}}$~=~0.55. We observe a similar trend for the variation in the total CARS intensity although the overall CARS intensity is reduced due to the lower collection angle. The effect of the detection NA is emphasized in Fig.~\ref{fig:Figure5}(c), which displays the ratio of the CARS intensity obtained with  NA$_{\mathrm{det}} = 0.55\/$ divided by that obtained using NA$_{\mathrm{det}} =0.95\/$. The color code in the ratio images has been scaled relative to the non-scattering case, where the white color corresponds to the ratio obtained when no scatterer is present. Blue regions indicate positions where the CARS signal ratio using these two detection NA's is smaller as compared to the non-scattering case. For majority of scatterer locations, the relative amount of signal loss due to scattering is greater for the lower detection of 0.55. Omission of large angles that have more signal contribution as shown in Fig.~\ref{fig:Figure4}(b)) increases the relative signal loss. Red areas, however, correspond to scatterer positions where higher ratios are observed. These latter regions tend to be in the shadow of the geometrical focus. The higher ratios occur due to scattering-induced redirection of excitation field density to areas that are otherwise depleted of excitation energy.

%%%%%%%%%%%% C %%%%%%%%%%%%%%%%%%%%
\subsection{Effect of incident polarization on CARS imaging in scattering samples}

In the previous Sections, we examined the variation of the CARS emission profiles and total signal as a function of scatterer position. These results show that both the angular distribution and the intensity of the CARS signal depend on the scatterer size and location. Here, we examine the effects of scatterers on CARS \textit{images}, and how these images are affected by the polarization state of the excitation beams.  

\begin{figure}[ht!]
	\begin{center}
		\includegraphics[width=0.5\columnwidth]{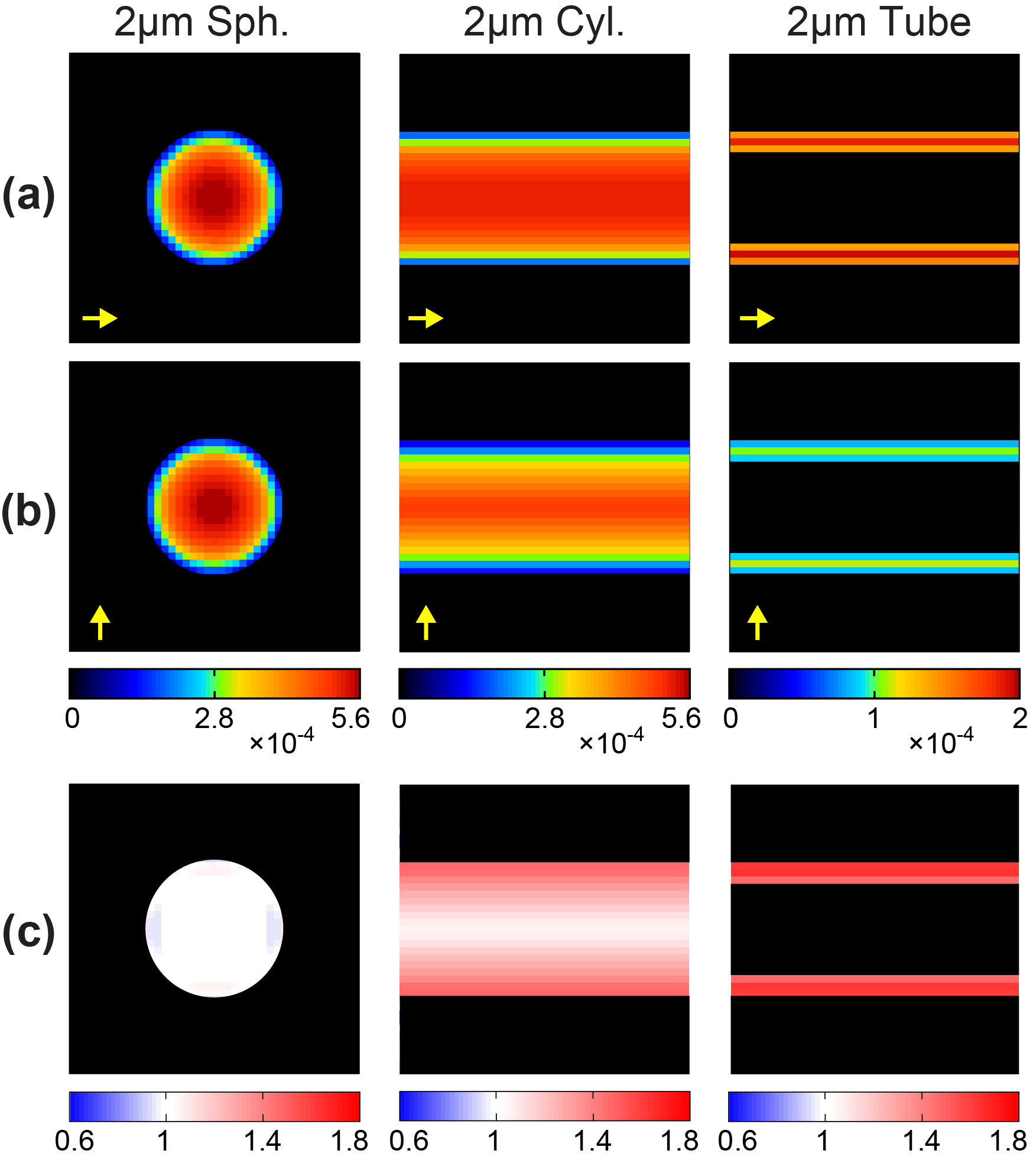}
	\end{center}
	\caption{CARS images ($x$-$y$ scan) of 2~$\mu$m sphere, 2~$\mu$m myelin cylinder,  and 2~$\mu$m myelin tube located at the focus for (a) $x\/$-polarized incident and (b) $y\/$-polarized incident upon the lens in a non-scattering medium. (c) The polarization ratio is calculated by dividing (a) by (b). Size of each image is 4.05$\mu\/$m~$\times$~4.05$\mu\/$m. Small arrows (yellow) show the orientation of the input polarization. Excitation and detection numerical apertures are fixed at NA$_{\mathrm{ex}}$~=~0.825 and NA$_{\mathrm{det}}$~=~0.95.
		\label{fig:Figure6}}
\end{figure} 

We first consider CARS images in the absence of scattering particles. In Fig.~\ref{fig:Figure6} we show simulated images of three objects: the 2~$\mu$m~diameter solid sphere, myelin cylinder and myelin tube considered previously. Figure \ref{fig:Figure6}(a) shows simulated CARS images obtained when both input beams are $x\/$-polarized, whereas Fig.~\ref{fig:Figure6}(b) provides images obtained using $y\/$-polarized incident beams. The differences in the images obtained using these different polarizations is a direct consequence of the anisotropy of $\chi^{(3)}$. Figure \ref{fig:Figure6}(c) displays the ratio of the $x\/$- and $y\/$-polarization images. The CARS ratio image of the sphere is uniform because an isotropic $\chi^{(3)}$ was chosen. We also see a ratio of 1 in the middle portions of the myelin cylinder. The outer boundaries of the myelin cylinder and myelin tube have polarization ratios larger than 1. Recall that the cylinder and myelin tube are modeled as radially-ordered lipid membrane sheets, which exhibit a highly anisotropic $\chi^{(3)}$ in the laboratory frame that changes with the orientation of the lipid. The non-uniform ratio images thus reflect the variation of the lipid orientation. 

\begin{figure}[hbt!]
	\begin{center}
		\includegraphics[width=0.95\columnwidth]{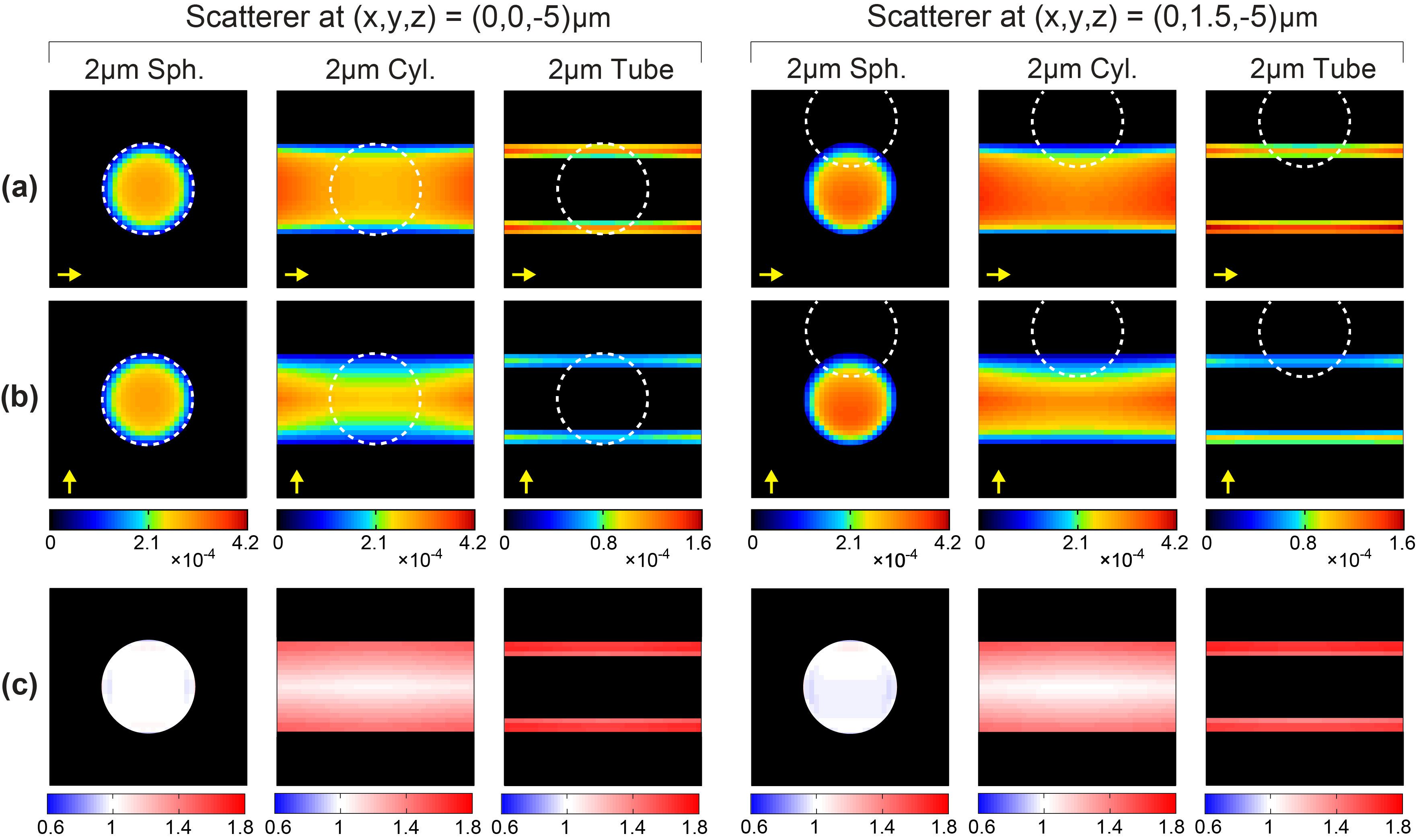}
	\end{center}
	\caption{CARS images ($x$-$y$ scan) of 2~$\mu$m sphere, 2~$\mu$m myelin-type cylinder, and 2~$\mu$m myelin tube located at the focus for (a) $x\/$-polarized incident and (b) $y\/$-polarized incident upon the lens. The results are shown for a spherical scatterer placed at $(x,y,z)=(0,0,-5)\,\mu$m (left) and $(x,y,z)=(0,1.5,-5)\,\mu$m (right). Dashed circle (white) shows the location of the scatterer. (c) The polarization ratio is calculated by dividing (a) by (b). Size of each image is 4.05$\mu\/$m~$\times$~4.05$\mu\/$m. Small arrows (yellow) show the orientation of the input polarization.  Excitation and detection numerical apertures are fixed at NA$_{\mathrm{ex}}$~=~0.825 and NA$_{\mathrm{det}}$~=~0.95.
		\label{fig:Figure7}}
\end{figure} 

Figure~\ref{fig:Figure7} shows simulated CARS images of the same objects in the presence of a scatterer at different locations. The scattering scenarios considered here correspond to those shown in Figs.~\ref{fig:Figure2}(c) and ~\ref{fig:Figure2}(d). Figs.~\ref{fig:Figure7}(a) and~\ref{fig:Figure7}(d) provide images obtained using $x\/$-polarized incident beams and Figs.~\ref{fig:Figure7}(b) and~\ref{fig:Figure7}(e) provide images obtained using $y\/$-polarized incident beams. Fig.~\ref{fig:Figure7}(c) and~\ref{fig:Figure7}(f) provide the polarization ratio images. In both scattering scenarios, the images are significantly distorted by the presence of the scatterer. For scatterers positioned along the optical axis, the center of the image is attenuated as a direct consequence of the scatterer position as seen in Fig.~\ref{fig:Figure4}. This is clearly seen for the myelin cylinder and myelin tube, whereas the scattering induced uniform amplitude attenuation is observed for the sphere. Once the scattering particle is offset by 1.5~$\mu$m relative to the optical axis, the image is distorted in an asymmetric pattern. Comparison of the two scenarios clearly demonstrate that scattering-induced distortions in CARS images are strongly affected by scatterer position. Moreover, the $x\/$-polarized and $y\/$-polarized images are affected somewhat differently by the scatterer. This difference originates primarily from the anisotropy of the lipid sample, which was also evident in Fig.~\ref{fig:Figure6}. As a result, the polarization ratio images do show sensitivity to the presence of the scatterer. The polarization ratio images shown in Figs.~\ref{fig:Figure7}(c) and~\ref{fig:Figure7}(f) differ by less than 5\% from the ratio images simulated in an non-scattering medium (Fig.~\ref{fig:Figure6}(c)). This is a consequence of the invariance of the scattered field from spherical particles relative to the linear polarization direction of the excitation field. Thus, while the CARS intensity image is affected significantly by the presence of scatterers, the polarization ratio image, which displays the anisotropy of the dipolar Raman scatterers, provides a much less distorted view of scattering samples.

Because of the complexity of the problem, we have avoided linear scattering by using an index-matched object in the focal volume. It is known that refractive index mismatches between objects and the surrounding medium within the focal volume give rise to additional effects~\cite{Djaker2006}. In particular, the internal and external scattered fields from refractive index mismatched objects slightly alter the electric fields in the $\mathbf{P}^{(3)}(\mathbf{r})$ calculation. In addition, the phase and amplitude differences between vibrationally resonant particles and the nonresonant medium can modify the CARS radiation in the far-field.~\cite{Cheng2013,Popov2012,Barlow2013,vanderKolk2016} However, including such additional effects complicates the analysis of wavefront distortions introduced by scattering particles in out-of-focus regions. Therefore, to isolate the effects of linear light scattering away from the focal volume, we have chosen to only consider non-resonant targets in a refractive index-matched medium. In future studies, we will examine index-mismatched objects in focus, as well as the effects of linear scattering from particles with more complicated geometrical shapes that are more representative of actual tissue structures.

% Beyond CARS microscopy, we have shown that HF-WEFS is a powerful approach for modeling incident beam propagation, focal field formation, signal generation, and emission in media containing scatterers. These capabilities provide opportunities for modeling scattering effect in other nonlinear optical microscopy techniques including second-harmonic generation (SHG), third-harmonic generation (THG) and two-photon excited fluorescence microscopy. 

%%%%%%%%%%%%%%%%%%%%%% Conclusion %%%%%%%%%%%%%%%%%%%%%%%%%%%
\section{Conclusion}
We have presented a computational framework to simulate the propagation, generation, and detection of CARS signals in a medium containing scattering particles at deterministic locations. We utilize the HF-WEFS method to simulate the pump and Stokes excitation fields, which are distorted by the presence of scatterers in the propagation path toward the focal region. Using the perturbed excitation fields, we calculate the spatially-dependent third-order polarization of objects in the focal volume, and apply the radiating dipole approximation to calculate the far-field CARS radiation. This framework is applied to examine the effects of scattering on the far-field CARS radiation pattern from three index-matched objects for various scatterer sizes and locations under different illumination and detection conditions.

Our results demonstrate that the presence of small scattering objects proximal to the focal volume results in an attenuation of the CARS signal intensity. The signal attenuation can be minimized by using lenses with increased NA for both excitation and detection. In addition to signal attenuation, we have also observed significant distortions to the angular distribution of the CARS radiation. This effect can be related to the scattering-induced phase shifts imprinted by the excitation fields on the nonlinear polarization in the focal volume. The attenuation and propagation direction of CARS radiation is highly dependent on the size and position of the particle, an observation that underlines that the effects of light-scattering in a coherent technique like CARS microscopy are complex and require a full view of amplitude and phase distortions. Our computations confirm that placement of a scattering object near focus produces noticeable artifacts in the CARS intensity image. However, our simulations also show that CARS anisotropy images are much less sensitive to the presence of spherical scatterers.

The framework and results presented in this work provide a platform for detailed mechanistic study of the effects of light scattering on the quality of CARS images. With subsequent improvements of the model, including the consideration of multiple scatterers and scatterers of varying shape and refractive properties, we expect that the bottom-up understanding gleaned from these simulations will foster the development of adaptive optics strategies for coherent nonlinear optical microscopy.

%%%%%%%%%%%%%%%%%%%%%% Acknowledgments%%%%%%%%%%%%%%%%%%%%%%%%%%%
\section*{Acknowledgments}
JCR, EOP and VV acknowledge support from the Laser Microbeam and Medical Program (LAMMP) a NIH Biomedical Technology Resource (P41-EB015890). EOP acknowledges support from the National Science Foundation (NSF CHE-1454885).


\begin{thebibliography}{99}
\bibitem{Cheng2013} J.-X. Cheng and X. S. Xie, \textit{Coherent Raman Scattering Microscopy} (CRC Press, 2013).

\bibitem{Chung2013} C.-Y. Chung, J. Boik, and E. O. Potma, "Biomolecular Imaging with Coherent Nonlinear Vibrational Microscopy," Annual Rev. Phys. Chem. \textbf{64}, 77--99 (2013).

\bibitem{Zhang2015} C. Zhang, D. Zhang, and J.-X. Cheng, "Coherent Raman Scattering Microscopy in Biology and Medicine," Annual Rev. Biomed. Eng. \textbf{17}, 415--445 (2015).

\bibitem{maker1965study} P. Maker and R. Terhune, "Study of optical effects due to an induced polarization third order in the electric field strength," Phys. Rev. \textbf{137}(3A) A801--A818 (1965).

\bibitem{Helmchen2005} F. Helmchen and W. Denk, "Deep tissue two-photon microscopy," Nature Methods \textbf{2}(12), 932--940 (2005).

\bibitem{Mosk2012} A. P. P. Mosk, A. Lagendijk, G. Lerosey, and M. Fink, "Controlling waves in space and time for imaging and focusing in complex media," Nature Photonics \textbf{6}(5), 283--292 (2012).

\bibitem{DeAguiar2015} H. B. de Aguiar, P. Gasecka, and S. Brasselet, "Quantitative analysis of light scattering in polarization-resolved nonlinear microscopy," Opt. Express \textbf{23}(7), 8960--8973 (2015).

\bibitem{Vellekoop2007} I. M. Vellekoop and A. P. Mosk, "Focusing coherent light through opaque strongly scattering media," Opt. Letters \textbf{32}(16), 2309--2311 (2007).

\bibitem{Yaqoob2008} Z. Yaqoob, D. Psaltis, M. S. Feld, and C. Yang, "Optical phase conjugation for turbidity suppression in biological samples," Nature Photonics \textbf{2}, 110-115 (2008).

\bibitem{Aguiar2016} H. B. de Aguiar, S. Gigan, and S. Brasselet, ``Enhanced nonlinear imaging through scattering media using transmission-matrix-based wave-front shaping,'' Phys. Rev. A {\bf94}, 043830 (2016).

\bibitem{Papadopoulos2016} I. N. Papadopoulos, J.-S Jouhanneau, J. F. A. Poulet and B. Judkewitz, "Scattering compensation by focus scanning holographic aberration probing (F-SHARP)," Nat. Photonics, \textbf{11}, 116--123 (2017).

\bibitem{Wright2007} A. J.Wright, S. P. Poland, J. M. Girkin, C. W. Freudiger, C. L. Evans, and X. S. Xie, "Adaptive optics for enhanced signal in CARS microscopy," Opt. Express \textbf{15}(26), 18209--18219 (2007).

\bibitem{Katz2014} O. Katz, E. Small, Y. Guan, and Y. Silberberg, "Noninvasive nonlinear focusing and imaging through strongly scattering turbid layers," Optica \textbf{1}(3), 170--174 (2014).

\bibitem{Cheng2002a} J.-X. Cheng, A. Volkmer, and X. S. Xie, "Theoretical and experimental characterization of coherent anti-Stokes Raman scattering microscopy," J. Opt. Soc. Am. B \textbf{19}(6), 1363-1375 (2002).

\bibitem{Zhu2013} C. Zhu and Q. Liu, "Review of Monte Carlo modeling of light transport in tissues," J. Biomed. Optics \textbf{18}(5) 050902 (2013).

\bibitem{Hokr2014} B. H. Hokr, V. V. Yakovlev, and M. O. Scully, "Efficient Time-Dependent Monte Carlo Simulations of Stimulated Raman Scattering in a Turbid Medium," ACS Photonics \textbf{1}(12), 1322--1329 (2014).

\bibitem{Lin2009} J. Lin, H. Wang, W. Zheng, F. Lu, C. Sheppard, and Z. Huang, "Numerical study of effects of light polarization, scatterer sizes and orientations on near-field coherent anti-Stokes Raman scattering microscopy," Opt. Express \textbf{17}(4), 2423--2434 (2009).

\bibitem{Lin2010} J. Lin, W. Zheng, H. Wang, and Z. Huang, "Effects of scatterers' sizes on near-field coherent anti-Stokes Raman scattering under tightly focused radially and linearly polarized light excitation," Opt. Express \textbf{18}(10), 10888--10895 (2010).

\bibitem{vanderKolk2016} J. van der Kolk, A. Lesina, and L. Ramunno, "Effects of refractive index mismatch on SRS and CARS microscopy," Opt. Express \textbf{24}(22), 25752--25766 (2016).

\bibitem{dunn2009OE} M. S. Starosta and A. K. Dunn, "Three-Dimensional Computation of Focused Beam Propagation through Multiple Biological Cells," Opt. Express \textbf{17}(15), 12455--12469 (2009).

\bibitem{Prahl2010a} S. A. Prahl, D. D. Duncan, and D. G. Fischer, "Monte Carlo propagation of spatial coherence," in Proc. SPIE, \textbf{7187} 71870G (2010).

\bibitem{Ranasinghesagara2014} J. C. Ranasinghesagara, C. K. Hayakawa, M. A. Davis, A. K. Dunn, E. O. Potma, and V. Venugopalan, "Rapid computation of the amplitude and phase of tightly focused optical fields distorted by scattering particles," J. Opt. Soc. Am. A \textbf{31}(7), 1520--1530 (2014).

\bibitem{Krishnamachari2007} V. V. Krishnamachari and E. O. Potma, "Focus-engineered coherent anti-Stokes Raman scattering microscopy: a numerical investigation," J. Opt. Soc. Am. A \textbf{24}(4), 1138--1147 (2007).

\bibitem{wang2005coherent} H. Wang, Y. Fu, P. Zickmund, R. Shi, and J.-X. Cheng, "Coherent anti-stokes Raman scattering imaging of axonal myelin in live spinal tissues," Biophys J. \textbf{89}, 581--591 (2005).

\bibitem{canta2016age} A. Canta, A. Chiorazzi, V. Carozzi, C. Meregalli, N. Oggioni, M. Bossi, V. Rodriguez-Menendez, F. Avezza, L. Crippa, R. Lombardi, G. de Vito, V. Piazza, G. Cavaletti, P. Marmiroli, "Age-Related Changes in the Function and Structure of the Peripheral Sensory Pathway in MiceNeurobiology of Aging," Neurobio. of Aging, \textbf{45} 136--148 (2016).

\bibitem{fu2011paranodal} Y. Fu, T. J. Frederick, T. B. Hu, G. E. Goings, S. D. Miller, and J.-X. Cheng, "Paranodal myelin retraction in relapsing experimental autoimmune encephalomyelitis visualized by coherent anti-Stokes Raman scattering microscopy," J. Biomed. Optics \textbf{16}(10), 106006 (2011).

\bibitem{imitola2011multimodal} J. Imitola, D. Cote, S. Rasmussen, X. S. Xie, Y. Liu, T. Chitnis, R. L. Sidman, C. P. Lin, and S. J. Khoury, "Multimodal coherent anti-Stokes Raman scattering microscopy reveals microglia-associated myelin and axonal dysfunction in multiple sclerosis-like lesions in mice," J. Biomed. Optics \textbf{16}, 021109 (2011).

\bibitem{deVito2016rp} G. de Vito, V. Cappello, I. Tonazzini, M. Cecchini, and V. Piazza, "RP-CARS reveals molecular spatial order anomalies in myelin of an animal model of Krabbe disease," J. Biophoton., 1-9 (2016).

\bibitem{shi2011longitudinal} Y. Shi, D. Zhang, T. B. Hu, X. Wang, R. Shi, X.-M. Xu, and J.-X. Cheng, "Longitudinal in vivo coherent anti-Stokes Raman scattering imaging of demyelination and remyelination in injured spinal cord," J. Biomed. Optics \textbf{16}, 106012 (2011).

\bibitem{Novotny2006} L. Novotny and B. Hecht, \textit{Principles of Nano-Optics} (Cambridge University, 2006).

\bibitem{Koay2011} C. G. Koay, "A simple scheme for generating nearly uniform distribution of antipodally symmetric points on the unit sphere," J. Comp. Sci. \textbf{2}(4), 377--381 (2011).

\bibitem{Richards1959} B. Richards and E. Wolf, "Electromagnetic diffraction in optical systems. II. Structure of the image field in an aplanatic system," in \textit{Proceedings of the Royal Society of London A} \textbf{253}(1274), 358--379 (1959).

\bibitem{VandeHulst1957} H. C. van de Hulst, \textit{Light scattering by small particles} (John Wiley \& Sons Inc, 1957).

\bibitem{Boyd2003} R. W. Boyd, \textit{Nonlinear Optics} (Academic Press, 2003).

\bibitem{deVito2012} G. de Vito, A. Bifone, and V. Piazza, "Rotating-polarization CARS microscopy: combining chemical and molecular orientation sensitivity," Opt. Express \textbf{20}(28), 29369--29377 (2012).

\bibitem{Mazely1987} T. L. Mazely and W. M. Hetherington, "Third-order susceptibility tensors of partially ordered systems," J. Chem. Phys. \textbf{87}(4), 1962--1966 (1987).

\bibitem{Belanger2009} E. B{\'{e}}langer, S. B{\'{e}}gin, S. Laffray, Y. {De Koninck}, R. Vall{\'{e}}e, D. C{\^{o}}t{\'{e}}, "Quantitative myelin imaging with coherent anti-Stokes Raman scattering microscopy: alleviating the excitation polarization dependence with circularly polarized laser beams," Opt. Express \textbf{17}(21), 18419--18432 (2009).

\bibitem{deVito2014rp} G. de Vito, I. Tonazzini, M. Cecchini, and V. Piazza, "RP-CARS: label-free optical readout of the myelin intrinsic healthiness," Opt. Express \textbf{22}(11), 13733--13743 (2014).

\bibitem{Jackson1975} J. D. Jackson, \textit{Classical Electrodynamics} (John Wiley \& Sons, Inc., 1975).

\bibitem{Djaker2006} N. Djaker, D. Gachet, N. Sandeau, P. F. Lenne, and H. Rigneault, "Refractive effects in coherent anti-Stokes Raman scattering microscopy," Applied Optics, \textbf{45}(27), 7005--7011 (2006).

\bibitem{Popov2012} K. I. Popov, A. F. Pegoraro, A. Stolow and L. Ramunno, "Image formation in CARS and SRS: effect of an inhomogeneous nonresonant background medium," Opt. Letters, \textbf{37}(4), 473--475 (2012).

\bibitem{Barlow2013} A. M. Barlow, K. Popov, M. Andreana, D. J. Moffatt, A. Ridsdale, A. D. Slepkov, J. L. Harden, L. Ramunno and A. Stolow, "Spatial-spectral coupling in coherent anti-Stokes Raman scattering microscopy," Opt. Express, \textbf{21}(13), 15298--15307 (2013).

\end{thebibliography}
\end{document}